\documentstyle[iopconf1]{article}
\newcommand{\beq}{\begin{equation}}
\newcommand{\eeq}{\end{equation}}
\def\bea{\begin{eqnarray}}
\def\eea{\end{eqnarray}}

\def\sss{\scriptscriptstyle}
\def\dbeta{\beta\beta_{0\nu}}
\def\eeWW{e^-e^-\to W^-W^-}
\def\ggmmww{\gamma\gamma \to \mu^+ \mu^+ W^- W^-}

\def\egnmmw{e^- \gamma \to \nu_e \mu^- \mu^- W^+}
\newcommand{\mw}{M_{\sss W}}
\def\gf{G_{\sss F}}
\def\rffly#1{\mathrel{\raise.3ex\hbox{$#1$\kern-.75em\lower1ex\hbox{$\sim$}}}}
\def\lsim{\rffly<}
\def\gsim{\rffly>}
\def\npb#1#2#3{19#2 {\em Nucl.\ Phys.} {\bf B#1} #3}
\def\plb#1#2#3{19#2 {\em Phys.\ Lett.} {\bf #1B} #3}
\def\prd#1#2#3{19#2 {\em Phys.\ Rev.} {\bf #1D} #3}
\newread\epsffilein 
\newif\ifepsffileok 
\newif\ifepsfbbfound 
\newif\ifepsfverbose 
\newdimen\epsfxsize 
\newdimen\epsfysize 
\newdimen\epsftsize 
\newdimen\epsfrsize 
\newdimen\epsftmp 
\newdimen\pspoints 
\def\epsfbox#1{\global\def\epsfllx{72}\global\def\epsflly{72}%
 \global\def\epsfurx{540}\global\def\epsfury{720}%
 \def\lbracket{[}\def\testit{#1}\ifx\testit\lbracket
 \let\next=\epsfgetlitbb\else\let\next=\epsfnormal\fi\next{#1}}%
\def\epsfgetlitbb#1#2 #3 #4 #5]#6{\epsfgrab #2 #3 #4 #5 .\\%
 \epsfsetgraph{#6}}%
\def\epsfnormal#1{\epsfgetbb{#1}\epsfsetgraph{#1}}%
\def\epsfgetbb#1{%
\openin\epsffilein=#1
\ifeof\epsffilein\errmessage{I couldn't open #1, will ignore it}\else
 {\epsffileoktrue \chardef\other=12
 \def\do##1{\catcode`##1=\other}\dospecials \catcode`\ =10
 \loop
 \read\epsffilein to \epsffileline
 \ifeof\epsffilein\epsffileokfalse\else
 \expandafter\epsfaux\epsffileline:. \\%
 \fi
 \ifepsffileok\repeat
 \ifepsfbbfound\else
 \ifepsfverbose\message{No bounding box comment in #1; using defaults}\fi\fi
 }\closein\epsffilein\fi}%
\def\epsfclipstring{}
\def\epsfsetgraph#1{%
 \epsfrsize=\epsfury\pspoints
 \advance\epsfrsize by-\epsflly\pspoints
 \epsftsize=\epsfurx\pspoints
 \advance\epsftsize by-\epsfllx\pspoints
 \epsfxsize\epsfsize\epsftsize\epsfrsize
 \ifnum\epsfxsize=0 \ifnum\epsfysize=0
 \epsfxsize=\epsftsize \epsfysize=\epsfrsize
 \epsfrsize=0pt
 \else\epsftmp=\epsftsize \divide\epsftmp\epsfrsize
 \epsfxsize=\epsfysize \multiply\epsfxsize\epsftmp
 \multiply\epsftmp\epsfrsize \advance\epsftsize-\epsftmp
 \epsftmp=\epsfysize
 \loop \advance\epsftsize\epsftsize \divide\epsftmp 2
 \ifnum\epsftmp>0
 \ifnum\epsftsize<\epsfrsize\else
 \advance\epsftsize-\epsfrsize \advance\epsfxsize\epsftmp \fi
 \repeat
 \epsfrsize=0pt
 \fi
 \else \ifnum\epsfysize=0
 \epsftmp=\epsfrsize \divide\epsftmp\epsftsize
 \epsfysize=\epsfxsize \multiply\epsfysize\epsftmp
 \multiply\epsftmp\epsftsize \advance\epsfrsize-\epsftmp
 \epsftmp=\epsfxsize
 \loop \advance\epsfrsize\epsfrsize \divide\epsftmp 2
 \ifnum\epsftmp>0
 \ifnum\epsfrsize<\epsftsize\else
 \advance\epsfrsize-\epsftsize \advance\epsfysize\epsftmp \fi
 \repeat
 \epsfrsize=0pt
 \else
 \epsfrsize=\epsfysize
 \fi
 \fi
 \ifepsfverbose\message{#1: width=\the\epsfxsize, height=\the\epsfysize}\fi
 \epsftmp=10\epsfxsize \divide\epsftmp\pspoints
 \vbox to\epsfysize{\vfil\hbox to\epsfxsize{%
 \ifnum\epsfrsize=0\relax
 \includegraphics{#1}%
 \else
 \epsfrsize=10\epsfysize \divide\epsfrsize\pspoints
 \includegraphics{#1}%
 \fi
 \hfil}}%
\global\epsfxsize=0pt\global\epsfysize=0pt}%
 {\catcode`\%=12 \global\let\epsfpercent=
\long\def\epsfaux#1#2:#3\\{\ifx#1\epsfpercent
 \def\testit{#2}\ifx\testit\epsfbblit
 \epsfgrab #3 . . . \\%
 \epsffileokfalse
 \global\epsfbbfoundtrue
 \fi\else\ifx#1\par\else\epsffileokfalse\fi\fi}%
\def\epsfempty{}%
\def\epsfgrab #1 #2 #3 #4 #5\\{%
\global\def\epsfllx{#1}\ifx\epsfllx\epsfempty
 \epsfgrab #2 #3 #4 #5 .\\\else
 \global\def\epsflly{#2}%
 \global\def\epsfurx{#3}\global\def\epsfury{#4}\fi}%
\def\epsfsize#1#2{\epsfxsize}
\begin{document}
\begin{flushright}
UdeM-GPP-TH-99-65
\end{flushright}

\title{Inverse neutrinoless double beta decay \\
  (and other $\Delta L=2$ processes)\footnote{Talk given at {\it
      Beyond the Desert -- Accelerator, Non-Accelerator and Space
      Approaches into the Next Millenium}, Castle Ringberg, Tegernsee,
    Germany, June 1999.}}

\author{David London\footnote{e-mail: london@lps.umontreal.ca}}

\affil{Laboratoire Ren\'e J.-A. L\'evesque, Universit\'e de
  Montr\'eal, C.P. 6128, succ.\ centre-ville, Montr\'eal, QC, Canada H3C 3J7}

\beginabstract 
I review the prospects for the detection of $\Delta L=2$ processes at
future colliders. Except in contrived models, the process $\eeWW$ is
unobservable at future linear colliders unless $\sqrt{s} \gsim 2$ TeV,
due to constraints from neutrinoless double beta decay. As there are
no analogous constraints on the Majorana mass of the $\nu_\mu$, $\mu^-
\mu^- \to W^- W^-$ could be observed at a muon collider with
considerably lower $\sqrt{s}$.  One can also consider esoteric
processes such as $\ggmmww$. Such processes may be observable if
$\sqrt{s} \gsim 4$ TeV.
\endabstract

There have been several talks at this conference dealing with
neutrinoless double beta decay ($\dbeta$). In this decay the
fundamental $\Delta L=2$ process is $W^- W^- \to e^- e^-$, mediated by
a Majorana neutrino. However, one can turn this around, and consider
$\eeWW$ \cite{invdbeta,BBLN,Gluza}. For obvious reasons, this process
is often referred to as ``inverse neutrinoless double beta decay,''
and it can in principle be explored at a future $e^+ e^-$ linear
collider (NLC) which is run in $e^-e^-$ mode.

\begin{figure}
\vskip +0.2truein
\centerline{\epsfxsize 4.0 truein \epsfbox {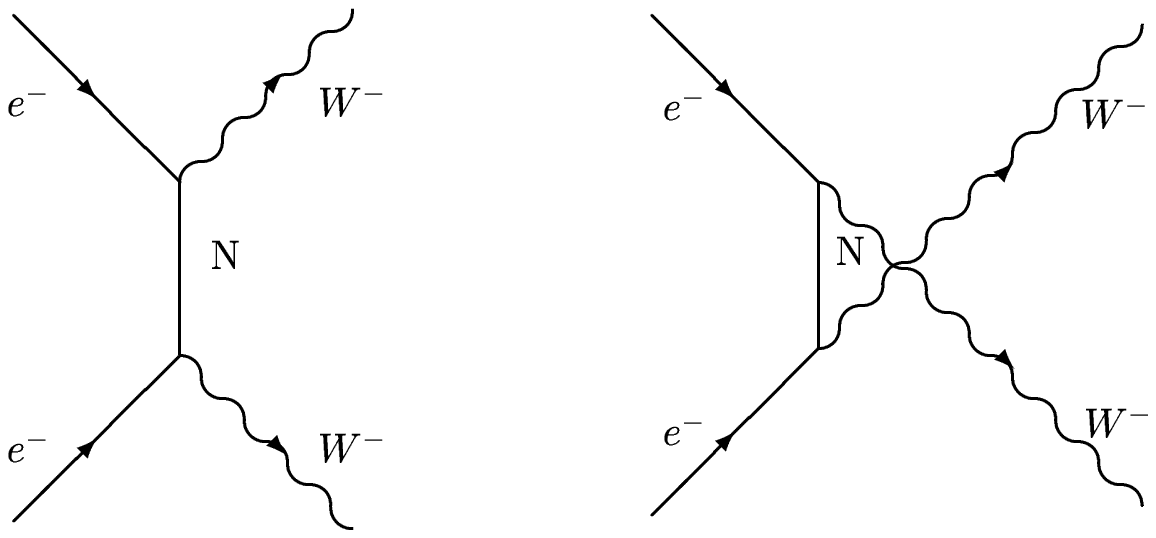}}
\vskip -4.2truein
\caption{Diagrams contributing to $\eeWW$.}
\label{eeWWdiags}
\end{figure}

The diagrams contributing to $\eeWW$ are shown in
Fig.~\ref{eeWWdiags}. The process is mediated by the exchange of a
(gauge-eigenstate) $\nu_e$. However, since the $\nu_e$ can mix with
other neutrinos,
\beq
\nu_e = \sum_i U_{ei} N_i ~,
\eeq
the mass eigenstates $N_i$ may be heavy or light, and all of these
states will contribute to $\eeWW$.

In the limit of $\sqrt{s} \gg \mw$, which is a good approximation for
the NLC, the cross section for $\eeWW$ can be written \cite{BBLN}
\beq
{d\sigma \over d\cos\theta} = {\gf^2\over 32\pi} 
\left( \sum_i M_i \left(U_{ei}\right)^2 
\left[ {t\over(t-M_i^2)} + {u\over(u-M_i^2)} \right] \right)^2 ~.
\label{Xsectionlong}
\eeq
There are two interesting limits to consider, which will be important
in what follows.
\begin{itemize}

\item $s \gg M_i^2$:
\beq
\sigma = {\gf^2\over 4\pi} 
\left( \sum_i M_i \left(U_{ei}\right)^2 \right)^2 ~.
\label{Xsectionunitarity}
\eeq

\item $M_i^2 \gg s$:
\beq
\sigma = {\gf^2\over 16\pi} s^2
\left( \sum_i {\left(U_{ei}\right)^2 \over M_i } \right)^2 ~.
\label{XsectionheavyN}
\eeq
\end{itemize}

From the above, we see that the two parameters entering into the
$\eeWW$ cross section are $\left(U_{ei}\right)^2$ and $M_i$. What are
the constraints on these quantities? First, the results from
low-energy experiments imply that \cite{Nardi}
\beq
\sum_{i\ne e} |U_{ei}|^2 \le 6.6 \times 10^{-3} ~.
\eeq

Second, we can consider unitarity. From Eq.~(\ref{Xsectionunitarity})
we see that the cross section does not vanish in the limit $\sqrt{s}
\to \infty$, violating unitarity. We therefore conclude that either
(a) 
\beq
\sum_i M_i \left(U_{ei}\right)^2 = 0 ~,
\label{massmixrelation}
\eeq
or (b) there are additional contributions to the process, such as the
$s$-channel exchange of a $\Delta^{--}$.  For the moment, let us
assume that there are no doubly-charged Higgs bosons. Is
Eq.~(\ref{massmixrelation}) satisfied? The answer to this question is
{\it yes}. It is straightforward to show that
\beq
\sum_i \left(U_{ei}\right)^2 M_i = M^*_{ee}~,
\eeq
where $M_{ee}$ is the Majorana mass of the $\nu_e$. Note that such a
mass can only arise in the presence of a Higgs triplet. However, since
we have assumed that there are no doubly-charged Higgses, this implies
that there are no Higgs triplets. We therefore conclude that $M_{ee} =
0$, so that Eq.~(\ref{massmixrelation}) is automatically satisfied.

As an explicit example, consider the well-known seesaw mechanism. One
adds a right-handed neutrino $N$ to the spectrum, and allows it to
acquire a Majorana mass via a Higgs singlet. The mass matrix then
looks like
\beq
\left(\matrix{0 & m \cr m & M \cr}\right),
\eeq
where $m$ is light ($m \sim m_e$) and $M$ is heavy. The two mass
eigenstates are $N_1$ and $N_2$, with masses $-m^2/M$ and $M$,
respectively. We can therefore write
\beq
\nu_e = N_1 \, \cos\theta + N_2 \, \sin\theta ~,
\eeq
where $\sin\theta = m/M$ ($\cos\theta \simeq 1$). With these masses
and mixings, one can clearly see that the relation in
Eq.~\ref{massmixrelation} is satisfied. I will return to this example
later, but one key point to retain is that the mixing naturally
satisfies $U_{ei} \sim m_e/M_i$. 

From the above discussion, we therefore conclude that unitarity puts
no constraints on $\left(U_{ei}\right)^2$ and $M_i$. However, there
are constraints from $\dbeta$. In $\dbeta$, the relevant scale is the
mass of the proton. For the $N_i$ which are light, $M_i \ll 1$ GeV,
measurements of $\dbeta$ restrict \cite{Vergados}
\beq
\langle m_\nu \rangle = \sum_{i~light} \left(U_{ei}\right)^2 M_i 
< 0.62~{\hbox{eV}}~.
\label{lightNlimit}
\eeq
On the other hand, for heavy neutrinos, $M_i \gg 1$ GeV, the relevant
quantity is $\sum_{i~heavy} \left(U_{ei}\right)^2 / M_i$, with the
constraint \cite{Vergados}
\beq
\sum_{i~heavy} \left(U_{ei}\right)^2 {1\over M_i} < 
1.0 \times 10^{-4}~{\hbox{TeV}}^{-1} ~.
\label{heavyNlimit}
\eeq

How does this all affect inverse $\dbeta$? Suppose, first, that all
the neutrinos are light. Combining the expression for the cross
section [Eq.~(\ref{Xsectionunitarity})] with the constraint in
Eq.~(\ref{lightNlimit}), we find
\beq
\sigma(\eeWW) \lsim 10^{-17}~fb~.
\label{tinyXsection}
\eeq
This is clearly unobservable. Let us now suppose that all the
neutrinos are heavy, i.e.\ $M_i \gg \sqrt{s}$. For $\sqrt{s} = 1$ TeV,
from Eqs.~(\ref{XsectionheavyN}) and (\ref{heavyNlimit}) we find
\beq
\sigma(\eeWW) <  0.01~fb~.
\eeq
For an NLC luminosity of 80 ${\rm fb}^{-1}$, this yields only 0.8
events/year, which is not promising. Furthermore, by the time the NLC
is built, the limits from $\dbeta$ will be much stronger. One must
therefore conclude that, due to low-energy constraints from $\dbeta$,
the process $\eeWW$ at a 1 TeV NLC will be unobservable.

But this raises a question: can the constraints from $\dbeta$ be
evaded? For example, consider the case of one heavy neutrino ($M \sim
1$ TeV) with mixing $U_{ei}^2 = 5 \times 10^{-3}$. Ignoring
$\dbeta$, this gives an $\eeWW$ cross section at $\sqrt{s} = 1$ TeV of
10 fb. This is enormous: it would yield 800 events/year.
Unfortunately, these mass and mixing parameters give $\left( U_{ei}
\right)^2 / M = 5 \times 10^{-3}~{\rm TeV}^{-1}$, in violation of the
constraints from $\dbeta$ [Eq.~(\ref{heavyNlimit})]. However, couldn't
we add other heavy neutrinos, whose factors of $\left( U_{ei}
\right)^2 / M$ cancelled this? The answer is obviously {\it yes}. For
example, we could add one neutrino of mass $M = 100$ GeV, with mixing
$U^2 = -5 \times 10^{-4}$. Or we could add 10 neutrinos of mass $M =
10$ TeV, with mixings $U^2 = -5 \times 10^{-3}$. In fact, there are an
infinite number of possibilities.

There are, however, a couple of problems with such scenarios. First,
they are all somewhat contrived, perhaps even fine-tuned. Second,
recall the seesaw mechanism I described earlier. Based on that
example, we would expect that $U_{ei} \sim m_e/M_i$. However, this is
not obeyed in the above scenarios: in these cases the mixing of the
heavier neutrinos is larger than, or equal to that of the lighter
neutrinos. What we conclude from these examples is that, although it
is possible to evade bounds from $\dbeta$ --- there are regions of
parameter space where cancellations occur \cite{Gluza} --- it does not
happen naturally. There is one more point to be made here: in the
scenario with a lighter neutrino of mass $M = 100$ GeV, the $N$ would
first be discovered in $e^+ e^- \to N \nu$, and we would be able to
verify that it is indeed a Majorana neutrino. Its mass and mixing
would be measured, and we would realize that it violated the $\dbeta$
bound. We would therefore {\it infer} the existence of heavier
neutrinos before they were discovered.

\begin{figure}
\vskip -0.2truein
\centerline{\epsfxsize 3.5 truein \epsfbox {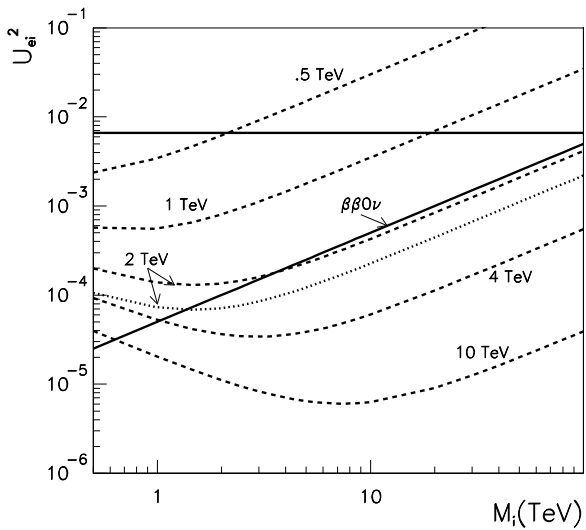}}
\vskip -0.4truein
\caption{Discovery limit for $\eeWW$ at the NLC 
  as a function of $M_i$ and $(U_{ei})^2$ for various values of
  $\protect\sqrt{s}$ (dashed lines). In all cases, the parameter space
  above the line corresponds to observable events. We also superimpose
  the experimental limit from $\dbeta$ (diagonal solid line), as well
  as the limit on $(U_{ei})^2$ (horizontal solid line).  Here, the
  parameter space above the line is ruled out.}
\label{eeWWXsection}
\end{figure}

To summarize the above discussion: it is possible to evade the
constraints from $\dbeta$, but one requires rather somewhat unnatural
scenarios. If we do not allow such solutions, we can calculate the
discovery limit for $\eeWW$ at the NLC, as a function of $U_{ei}^2$
and $M_i$, for various values of $\sqrt{s}$. The results are shown in
Fig.~\ref{eeWWXsection}, where we have assumed a luminosity of $80
(\protect\sqrt{s}/{\hbox{(1 TeV)}})^2~fb^{-1}$ and demanded 10 events
for discovery\footnote{In this figure, taken from Ref.~\cite{BBLN},
  the constraint from $\dbeta$ has been taken to be $\sum_{i~heavy}
  \left(U_{ei}\right)^2 {1\over M_i} < 0.7 \times
  10^{-4}~{\hbox{TeV}}^{-1}$, which is slightly stronger than that
  given in Eq.~(\ref{heavyNlimit}).}. From this figure we see that
constraints from $\dbeta$ essentially rule out the observation of the
process $\eeWW$ at an NLC of $\sqrt{s} = 2$ TeV or less. For a 4 TeV
or 10 TeV NLC, there exists a sizeable region of $M_i$-$(U_{ei})^2$
parameter space, not ruled out by $\dbeta$, which produces an
observable signal for $\eeWW$.

So far, the entire discussion has taken place assuming that there is
no doubly-charged Higgs boson. Let us now return to the beginning and
assume that there is a $\Delta^{--}$ which contributes to $\eeWW$. How
does this change things? In fact, it does not change the conclusions
at all. To see this, consider the left-right symmetric model. This
model contains a $\Delta^{--}$, which is a member of a representation in
which the neutral partner obtains a vacuum expectation value and gives
mass to both the $W$ and $\nu_e$. The $\Delta^{--}$ contribution to
$\eeWW$ involves the product of $g_{\Delta ee}$ and $g_{\Delta \sss WW}$,
which are the couplings of the $\Delta^{--}$ to $e^-e^-$ and $W^-W^-$,
respectively. However, direct calculation gives
\beq
g_{\Delta ee} g_{\Delta \sss WW} \propto g^2 {M_{ee} \over M_\Delta}~,
\eeq
where $M_{ee}$ is the Majorana mass of the $\nu_e$. However, from
Eq.~(\ref{lightNlimit}), we have $M_{ee} \lsim 1$ eV (since $U_{ee}
\simeq 1$), so that the product $g_{\Delta ee} g_{\Delta \sss WW}$ is
tiny.  Thus, the exchange of the $\Delta^{--}$, though necessary for
unitarity, does not affect the cross sections, and so the previous
conclusions hold. (In fact, if the seesaw mechanism is used, the
process $\eeWW$ is completely unobservable, as per
Eq.~(\ref{tinyXsection}).)

Since the prospects for the observation of $\eeWW$ are relatively
poor, the obvious next question is: what about other $\Delta L=2$
processes? The key observation is that, although $\dbeta$ strongly
constrains the Majorana mass of the $\nu_e$, $\nu_\mu$ (and
$\nu_\tau$) are not similarly constrained.  Therefore it stands to
reason that one should consider $\Delta L=2$ processes involving
$\nu_\mu$.

First, there has been much discussion recently about the prospects for
building a linear muon collider. If such a collider can be built, then
it can be run in $\mu^-\mu^-$ mode, and one can look for the process
$\mu^-\mu^- \to W^-W^-$. In this case, the only constraint is on
$\nu_\mu$ mixing \cite{Nardi}:
\beq
\sum_{i\ne \mu} |U_{\mu i}|^2 \le 6.0 \times 10^{-3} ~.
\eeq
The cross section is therefore quite sizeable, even at $\sqrt{s} =
500$ GeV. (To see this, simply refer to Fig.~\ref{eeWWXsection},
ignoring the constraints from $\dbeta$).

\begin{figure}
\vskip -0.2truein
\centerline{\epsfxsize 3.5 truein \epsfbox {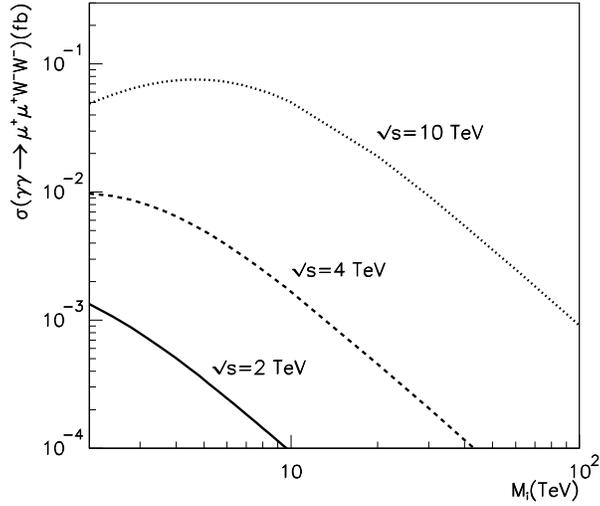}}
\vskip -0.4truein
\caption{Cross section for $\ggmmww$ at the NLC assuming 
  $(U_{\mu i})^2 = 6.0 \times 10^{-3}$ for $\protect\sqrt{s}=2$ TeV
  (solid line), 4 TeV (dashed line) and 10 TeV (dotted line).}
\label{ggllwwXsection}
\end{figure}

Second, one can consider more esoteric possibilities, such as
$\ggmmww$ \cite{BBLN}. The cross section for this process is shown in
Fig.~\ref{ggllwwXsection}. Assuming a luminosity of $80
(\sqrt{s}/{\hbox{(1 TeV)}})^2~fb^{-1}$, one sees from this figure that
$\ggmmww$ requires $\sqrt{s} \gsim 4$ TeV for observability. (Note,
however, that if $M_i < \sqrt{s}$, the new neutrino is far more likely
to be first discovered via single production in $e^+ e^- \to \nu_\mu
N_i$ than in $\ggmmww$.)

Finally, there are other, even more exotic processes, such as
$\egnmmw$ \cite{BBLN}. However, the cross sections for such processes
are even smaller than those for $\ggmmww$.

To conclude: if a linear $e^+ e^-$ collider is ever built, it can in
principle be run in $e^- e^-$ mode in order to search for the $\Delta
L=2$ process $\eeWW$. However, except in rather contrived scenarios,
constraints from low-energy neutrinoless double beta decay effectively
rule out this process unless $\sqrt{s} > 2$ TeV. Of course, by the
time such a collider is built, the constraints from $\dbeta$ will
probably be considerably more stringent, so that even higher
centre-of-mass energies will be required.

As there are no analogous constraints on the Majorana mass of the
$\nu_\mu$, one can consider $\Delta L = 2$ processes involving a
$\nu_\mu$. For example, should a muon collider be built, the process
$\mu^- \mu^- \to W^- W^-$ could be readily observable, even for
$\sqrt{s} = 500$ GeV. In the absence of such a collider, one must
consider more esoteric processes such as $\ggmmww$.  These are
observable for $\sqrt{s} \gsim 4$ TeV.

\section*{Acknowledgements}

I would like to thank G. B\'elanger, F. Boudjema and H. Nadeau, who
collaborated with me on the topics discussed here. This research was
financially supported by NSERC of Canada and FCAR du Qu\'ebec.

\end{document}